%% file: Draft_1.tex
\documentclass[a4paper,11pt]{article}
\pdfoutput=1 % if your are submitting a pdflatex (i.e. if you have
\usepackage{jheppub} % for details on the use of the package, please
\usepackage[T1]{fontenc} % if needed
\usepackage{comment}

\DeclareMathAlphabet{\mathpzc}{OT1}{pzc}{m}{it}

\input{def_jazs_tau.tex}

\begin{document}

\title{\boldmath Resonant $CP$ violation in rare $\tau^{\pm}$ decays}
%\title{Resonant CP violation in $\tau^{\pm}$ decays.}
\author{Jilberto Zamora-Saa}
\emailAdd{jzamorasaa@jinr.ru}
%\pacs{14.60.St,13.20-v,13.15.+g}

\affiliation{
Dzhelepov Laboratory of Nuclear Problems, Joint Institute for Nuclear Research, Dubna 141980, Russia.}

\abstract{
In this work, we study the lepton number violating tau decays  via intermediate on-shell Majorana neutrinos $N_j$ into two scalar mesons and a lepton $\tau^{\pm} \to M_1{}^{\pm} N_j \to M_1^{\pm} M_2^{\pm} \ell^{\mp}$. 
We calculate the Branching ratios $Br(\tau^{\pm})$ and the CP asymmetry $(\Gamma(\tau^+) - \Gamma(\tau^-))/(\Gamma(\tau^+) + \Gamma(\tau^-))$ for such decays, in a scenario that contains at least two heavy Majorana neutrinos. 
The results show that the CP asymmetry is small, but becomes comparable with
the branching ratio ${\rm Br}(\tau^{\pm})$ when their mass difference is similar with their decay width $\Delta M_N \sim  \Gamma_N$. We also present regions of the heavy-light neutrino mixing elements, in which the $CP$ asymmetry could be explored in future tau factories.}

\maketitle

%%%%%%%%%%%%%%%%%%%%%%%%%%%%%%%%%%%%%
%%%%%%%%%%%%%%%%%%%%%%%%%%%%%%%%%%%%%
\section{Introduction}
\label{intro}

During the last decades, neutrino experiments that have shown that neutrinos have non-zero masses~\cite{Fukuda:1998mi,Eguchi:2002dm}, also suggest that the first three mass eigenstates are very light with masses $\sim 1$ eV, and the mixing between flavour and mass eigenstates is characterized by the Pontecorvo-Maki-Nakagawa-Sakata Matrix, $U_\txty{PMNS}$~\cite{Maki:1962mu}. Therefore, if these light masses are produced by means of some see-saw mechanism \cite{Mohapatra:2005wg,Mohapatra:2006gs}, the existence of one or more  heavier neutrinos is needed. The current experimental uncertainties in the $B_\txty{PMNS}$ matrix elements allow introduce these new heavy neutral leptons called sterile neutrinos (SN)~\cite{Antusch:2008tz,Malinsky:2009gw,Dev:2009aw,Forero:2011pc,Das:2017nvm}, however the small values of these uncertainties imply a strongly suppressed interaction between standard model (SM) particles and SN.  In addition, due to the fact that neutrinos are massive particles,  a fundamental question arises: are neutrinos Dirac or Majorana particles?, If neutrinos are Dirac particles, the reactions in which they participate must preserve the lepton number ($\Delta L=0$). On the contrary, if neutrinos are Majorana particles, they are indistinguishable from their antiparticles, and the lepton number can be violated in two units ($\Delta L=2$). On the other hand, Neutrino oscillations (NOs) experiments have confirmed  that $\theta_{13}$ angle of $B_\txty{PMNS}$ is non zero~\cite{Beringer:1900zz,GonzalezGarcia:2012sz}, thus, the possibility of $CP$ violation in the light neutrino sector is still open; nevertheless, extra sources of $CP$ violation are needed in order to explain Baryogenesis via Leptogenesis \cite{Strumia:2006qk}. Recent studies explored the $CP$ violation and the phenomenology of SN neutrinos in the context of rare meson decays ~\cite{Dib:2000wm,Cvetic:2012hd,Cvetic:2013eza,Cvetic:2014nla,Cvetic:2015naa,Cvetic:2015ura,Dib:2014pga,Zamora-Saa:2016qlk}, however, in this work we will focus in the phenomenology of the rare tau decays \cite{Helo:2010cw,Gribanov:2001vv,Helo:2011yg} in the framework of tau factories, such as Super Charm-Tau Factory (CTF) in the Budker Institute of Nuclear Physics (Novosibirsk, Russia), \cite{Levichev:2008zz,Eidelman:2015wja} making it possible to extend the SN searches to tau decay processes. In this letter we focus in the rare decays of  tau leptons into two scalar mesons and one charged lepton ($\ell = e, \mu$),  via two on-shell intermediate neutrinos $N_j$, and look for the possibility of detection of CP asymmetries in such decays. The relevant processes are the lepton number violating channels  $\tau^{\pm} \to M_1^{\pm} M_2^{\pm} \ell^{\mp} $ where $M_1$,$M_2$$=\pi$,$K$ and $\ell=e$,$\mu$.  We also show that the branching ratios are very small\footnote{Both the branching ratio as $CP$ asymmetries are proportional to the  product of  square mixing elements $|B_{\tau N}|^2 |B_{\ell N}|^2$}, but  could be appreciable enough and could be measured in future $\tau$ factories where huge numbers of taus will be produced \cite{Bondar2013,Eidelman:2015wja}, if the heavy-light neutrino mixing elements are sufficiently large but still lower than the present upper bounds.

The program of this paper is the following: in Section~\ref{s1} we present the notation and formalism for the rare tau decay; in Sections~\ref{sBR} we present the relevant expression for the branching ratio calculations; in Sections~\ref{sCP} we present the relevant expression for the $CP$ asymmetries calculations; in Sections~\ref{sres} we present the results of the relevant parameters for the future searches; finally, in section~\ref{s_conclusion} we present the summary and conclusions.
%%%%%%%%%%%%%%%%%%%%%%%%%%%%%%%%%%%%%%%%%%%%%%%%%%%%%%%%%%%%%%%%%%%%%%%%%%%%%%%%%%%%%%%%%%%%%%%%%%%%%%%%%%%%%%%%%%%%%%%%%%%%%%%%%%%%%%%%%%%%%%%%%%%%%%%%%%%%%%%%%%%%%%%%%%%%%%%%%%%%%%%%%%%%%%%%%%%%

\section{Process and Formalism}
\label{s1}
\begin{figure}[H]
\includegraphics[scale = 0.19]{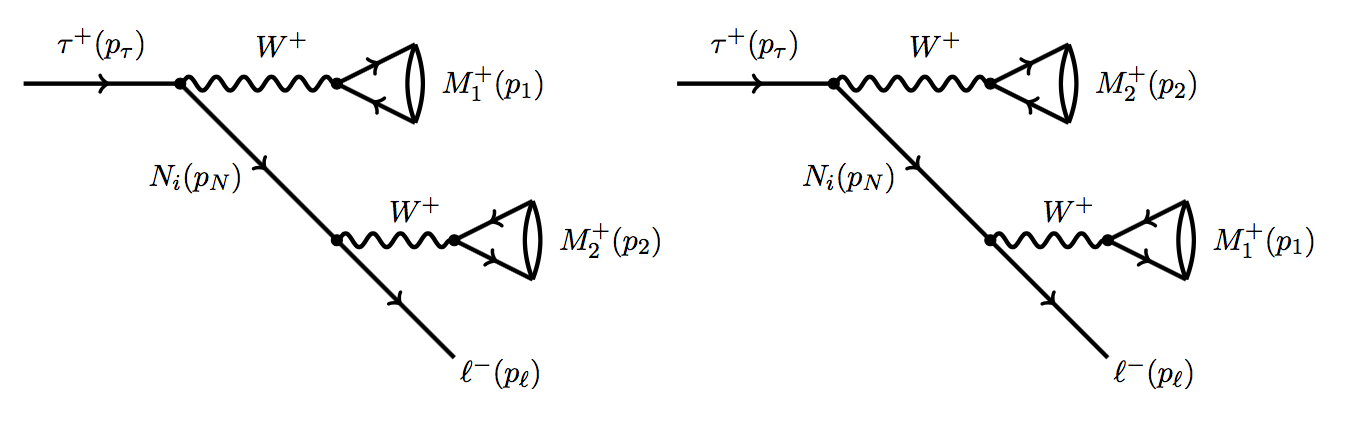}
\caption{Feynmann diagrams for the process $\tau^+\rightarrow M_1^+ M_2^+ \ell^-$. Left side: Direct channel $D$. Right side: Crossed channel $C$.}
\label{Fig1}
\end{figure}

\begin{figure}[H]
\includegraphics[scale = 0.19]{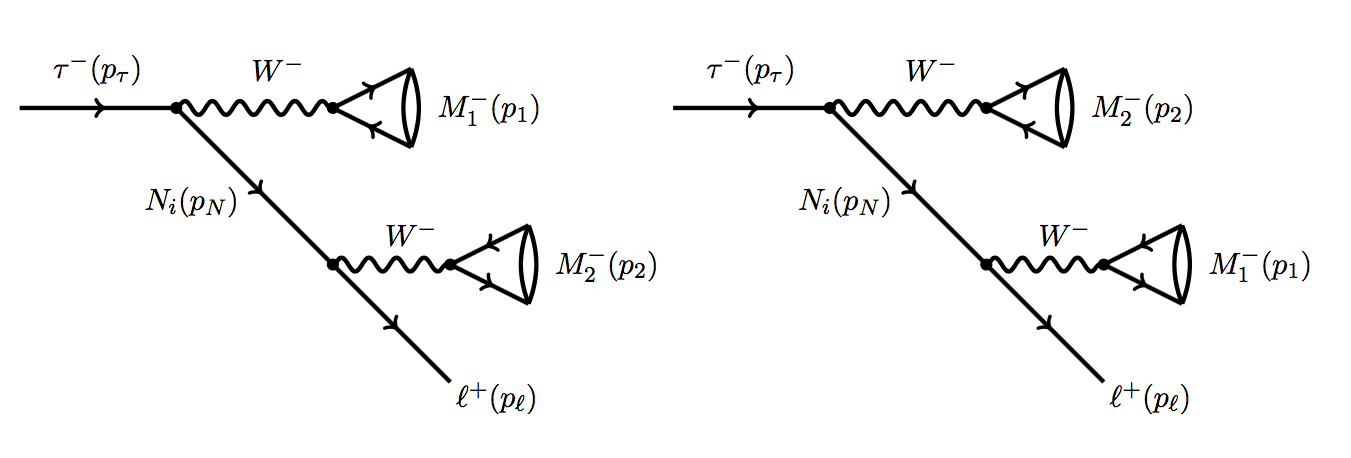}
\caption{Feynmann diagrams for the process $\tau^-\rightarrow M_1^- M_2^- \ell^+$. Left side: Direct channel $D$. Right side: Crossed channel $C$.}
\label{Fig2}
\end{figure}

As we stated above, we are interested in studying the $ \Delta L = 2 $ rare tau decays mediated by two on-shell heavy ($ 0.140 \le M_N \le 1.638$ GeV) Majorana neutrinos with the expectation of obtaining $CP$ violating signal in the neutrino sector. The relevant Feynman diagrams of the studied processes are presented in Fig.~\ref{Fig1} and Fig.~\ref{Fig2} for $\tau^+\rightarrow M_1^+ M_2^+ \ell^-$ and $\tau^-\rightarrow M_1^- M_2^- \ell^+$, respectively  

\begin{figure}[h]
\includegraphics[scale = 0.5]{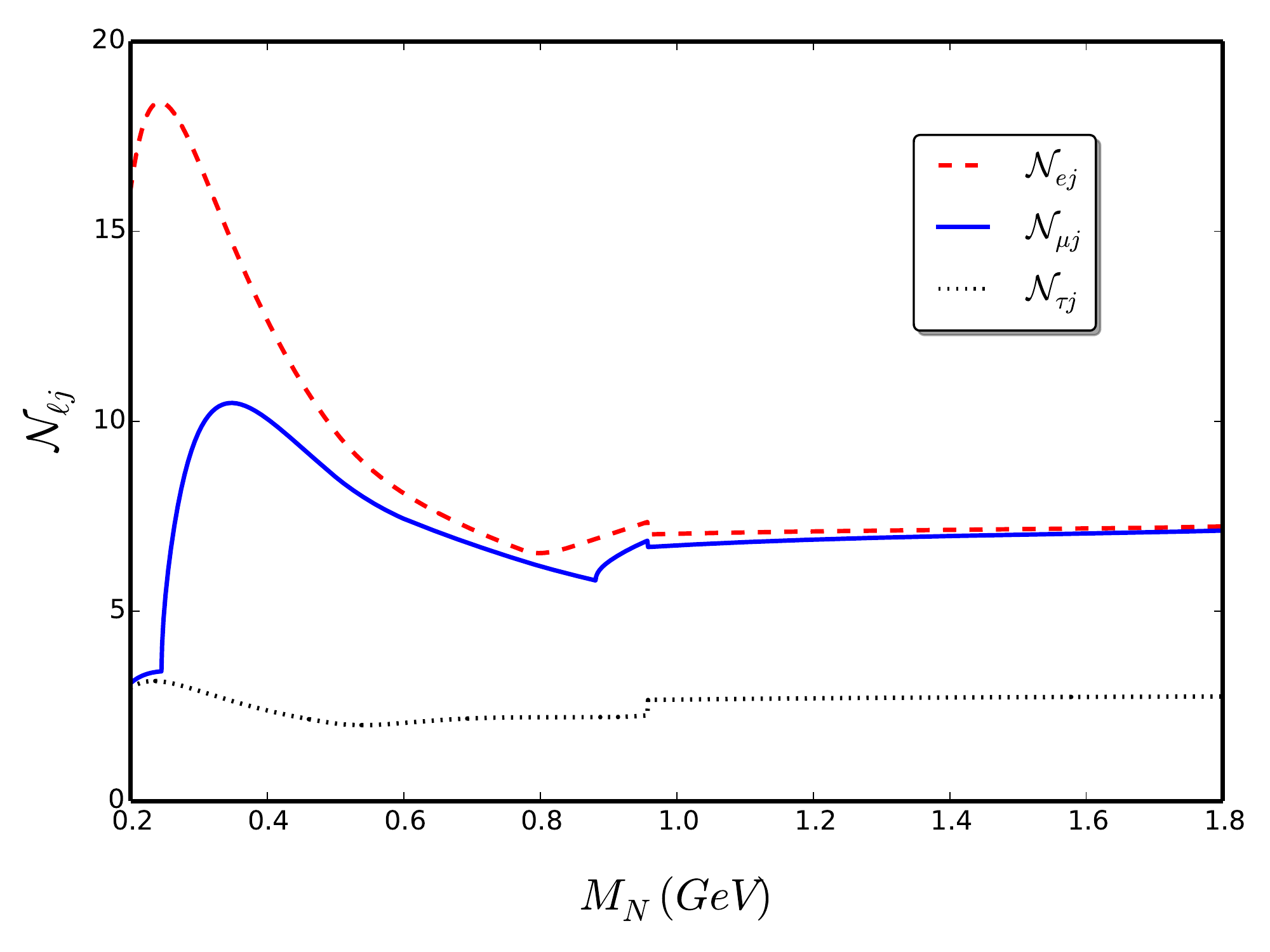}
\caption{Effective mixing coefficients. The dashed line (online red) is for ${\cal N}_{e j}$, solid line (online blue) for ${\cal N}_{\mu j}$ and the dotted one (online black) for ${\cal N}_{\tau j}$.}
\label{efcoef}
\end{figure}

In order to write down the amplitude and all the relevant quantities, we first define the neutrino flavor state as:
\be
\label{fstat}
\nu_{\ell}=\sum_{i=1}^3 B_{\ell i}\nu_i + \sum_{j=1}^n B_{\ell N_j} N_j \ ,
\ee
where $B_{\ell N_j}$ are the elements of the $PMNS$ matrix\footnote{Experimental limits for $|B_{\ell N_j}|^2$  in our mass range of interest are presented in figure Fig.~\ref{explim}.}  (heavy-light neutrino mixings elements) which are define as follow
\be
\label{PMNS}
 B_{\ell N_j} =  |B_{\ell N_j}| e^{i \phi_{\ell N_j}} \ ,
\ee
the left side of Eq.~(\ref{fstat}) stand for light neutrino sector and the right side for the heavy neutrino sector.
 The amplitude for a general process involving $n$ sterile neutrinos is\footnote{The definitions $\mathcal{M}^D_{\pm}$ and $\mathcal{M}^C_{\pm}$ can be understood as the amplitude for the direct channel and for the crossed one, respectively. Furthermore, the squared amplitude probability for the process will be  ${|\mathcal{M}_{\pm}|^2}={|\mathcal{M}_{\pm}^D|^2}+{|\mathcal{M}_{\pm}^C|^2} +  \mathcal{M}_{\pm}^D \mathcal{M}_{\pm}^{C \dagger}  + \mathcal{M}_{\pm}^{D \dagger} \mathcal{M}_{\pm}^C$ .}
 \begin{footnotesize}
\begin{subequations}
\begin{align}
\label{amp}
\nonumber i \mathcal{M_+} \equiv & \ i \mathcal{M}(\tau^+ \to M_1^+ M_2^+ \ell^- )=\mathcal{M}_{+}^D+\mathcal{M}_{+}^C =\\  &\underbrace{G_F^2 f_{M_1} f_{M_2}   V_{M_1} V_{M_2} B_{\ell N_j}^{}B_{\tau N_j}^{*}  P_{j}(D)\ \slashed{L}_+^D }_{\mathcal{M}_{+}^D} \ +\  \underbrace{G_F^2 f_{M_1} f_{M_2}   V_{M_1} V_{M_2}  B_{\ell N_j}^{} B_{\tau N_j}^{*}  P_{j}(C)\ \slashed{L}_+^C}_{\mathcal{M}_{+}^C} \ ,\\
\\ \nonumber
\nonumber i \mathcal{M_-} \equiv & \ i \mathcal{M}(\tau^- \to M_1^- M_2^- \ell^+ )=\mathcal{M}_{-}^D+\mathcal{M}_{-}^C =\\  & \underbrace{G_F^2 f_{M_1} f_{M_2}   V_{M_1}^{*} V_{M_2}^{*} B_{\ell N_j}^{*} B_{\tau N_j}^{}  P_{j}(D)\ \slashed{L}_-^D }_{\mathcal{M}_{-}^D} \ +\  \underbrace{G_F^2 f_{M_1} f_{M_2}   V_{M_1}^{*} V_{M_2}^{*}  B_{\ell N_j}^{*} B_{\tau N_j}^{} P_{j}(C) \slashed{L}_-^C}_{\mathcal{M}_{-}^C} \ , 
\end{align}
\end{subequations}
\end{footnotesize}
where $f_{1}$ and $f_{2}$ are the meson decay constants of $M_1^{\pm}$ and $M_2^{\pm}$,
and $V_{M_1}$, $V_{M_2}$ are the mixings elements of CKM matrix corresponding to mesons  $M_1$ and $M_2$, respectively.
The factors $\slashed{L}_{\pm}^D$ and $\slashed{L}_{\pm}^C$ contain the information related to the kinematics and are given by
\begin{align}
\slashed{L}_+^D&=\bar{u}(p_{\ell}) \slashed{p}_2 \slashed{p}_1 P_{j}(D) (1+\gamma_5) u(p_{\tau}) \quad ; \quad 
\slashed{L}_+^C=\bar{u}(p_{\ell}) \slashed{p}_1 \slashed{p}_2 P_{j}(C) (1+\gamma_5) u(p_{\tau})\ ,\\
\nonumber \\ 
\slashed{L}_-^D&=\bar{u}(p_{\ell}) \slashed{p}_2 \slashed{p}_1 P_{j}(D) (1+\gamma_5) u(p_{\tau}) \quad ; \quad 
\slashed{L}_-^C=\bar{u}(p_{\ell}) \slashed{p}_1 \slashed{p}_2 P_{j}(C) (1+\gamma_5) u(p_{\tau}) \ ,
\end{align}
and finally the factors $P_{j}(D)$ and $P_{j}(C)$ are the heavy Majorana neutrino propagators
\be
\label{propa}
P_{j}(D)=\sum_{j=1}^n  \frac{M_{N_j}}{(p_{\tau}-p_1)^2 -M_{N_j}^2+i \Gamma_{N_j} M_{N j}} \ ; \quad P_{j}(C)=\sum_{j=1}^n  \frac{M_{N_j}}{(p_{\tau}-p_2)^2 -M_{N_j}^2+i \Gamma_{N_j} M_{N j}}\ ,
\ee
here $\Gamma_{N_j}$ is the total decay width of the intermediate neutrinos, and can be approximated as follow
\begin{equation}
\Gamma_{N_j}  \approx  \K_j^{Ma}\ \frac{G_F^2 M_{N_j}^5}{96\pi^3} \, ,
\label{DNwidth}
\end{equation}
where
\begin{equation}
\K_j^{Ma} \equiv \K_j^{}(M_{N_j})  = {\cal N}_{e j}^{} \; |B_{e N_j}|^2 + {\cal N}_{\mu j}^{} \; |B_{\mu N_j}|^2 + {\cal N}_{\tau j}^{} \; |B_{\tau N_j}|^2
\, ,
\label{calK}
\end{equation}
the factors ${\cal N}_{\ell j}^{}$ being effective mixing coefficients and are presented in Fig.~\ref{efcoef} for our mass range of interest.

The decay with of the process is given as follow
\begin{align}
\label{dw}
\Gamma(\tau^{\pm} \rightarrow M_1^{\pm} M_2^{\pm} \ell^{\mp}) &\equiv \Gamma(\tau^{\pm}) =  \frac{1}{2!}(2 - \delta_{M_1 M_2}) \frac{1}{2 M_{\tau}}\int \overline{|\mathcal{M}_{\pm}|^2} \ d_3 \quad,
\end{align}
where  $\frac{1}{2!}(2 - \delta_{M_1 M_2})$ is the symmetry factor that counts for identical particles in the final states, $d_3$ denotes the number of states available per unit of energy in the 3-body final state\footnote{The decomposition of the 3-body phase space is presented in Appendix \ref{PS}.}.
\be
d_3 \equiv \frac{d^3 {\vec p}_1}{2 E_{1}({\vec p}_1)} 
 \frac{d^3 {\vec p}_2}{2 E_{2}({\vec p}_2)}
 \frac{d^3 {\vec p}_{\ell}}{2 E_{\ell}({\vec p}_{\ell})}
\delta^{(4)} \left( p_{\tau} - p_1 - p_2 - p_{\ell} \right) \ ,
\label{d3M}
\ee
here, $p_1$ and $p_2$ denote the momenta of $M_1$ and $M_2$ respectively, and $p_{\ell}$ the momentum of the charged lepton (see Fig.~\ref{Fig1} and Fig.~\ref{Fig2}).

%%%%%%%%%%%%%%%%%%%%%%%%%%%%%%%%%%%%%%%%%%%%%%%%%%%%%%%%
%%%%%%%%%%%%%%%%%%%%%%%%%%%%%%%%%%%%%%%%%%%%%%%%%%%%%%%%
%%%%%%%%%%%%%%%%%%%%%%%%%%%%%%%%%%%%%%%%%%%%%%%%%%%%%%%%
%%%%%%%%%%%%%%%%%%%%%%%%%%%%%%%%%%%%%%%%%%%%%%%%%%%%%%%%

\section{Branching ratio of $\tau^{\pm} \to M_1^{\pm} N_j \to M_1^{\pm} M_2^{\pm} \ell^{\mp}$ decays}
\label{sBR}

In a scenario with $n=2$ sterile neutrinos, the decay widths presented in Eq.(\ref{dw}) can be written as the double sum
of the contributions of $N_i$ and $N_j$  ($i,j=1,2$), with the mixing elements factored out
\begin{align}
%ec.(83) of the notes
\nonumber \Gamma(\tau^{\pm}) = &  \frac{1}{2!}(2 - \delta_{M_1 M_2})
\sum_{i=1}^2 \sum_{j=1}^2 k_i^{(\pm)} k_j^{(\pm)*} 
\\
& \times {\big [}
\widetilde{\Gamma}_{\tau}(DD^{*})_{ij} + \widetilde{\Gamma}_{\tau}(CC^{*})_{ij}+ \widetilde{\Gamma}_{\tau \pm}(DC^{*})_{ij} + \widetilde{\Gamma}_{\tau \pm}(CD^{*})_{ij} {\big ]} \ ,
\label{dwp}
\end{align}
here $\widetilde{\Gamma}$'s are the canonical decay widths (without heavy-light explicit mixing), and $k_j^{(\pm)}$ are parameters which contain the corresponding mixing factors and are presented in Eq.~(\ref{MF}). 
\be
k_j^{(+)} = B_{\ell N_j}^{} B_{\tau N_j}^{*} \ ,\qquad  k_j^{(-)}= (k_j^{(+)})^{*} \ .
\label{MF}
\ee
Due to the fact that $|\slashed{L}_+^D|^2 = |\slashed{L}_-^D|^2$ and  $|\slashed{L}_+^C|^2 = |\slashed{L}_-^C|^2$, we can omit the subscripts $\pm$ in the contribution terms $\widetilde{\Gamma}_{\tau}(DD^{*})_{ij}$ and $\widetilde{\Gamma}_{\tau}(CC^{*})_{ij}$ in Eq.~(\ref{dwp}). The canonical decay widths $\widetilde{\Gamma}_{\tau \pm}(XY^{*})_{ij}$, where $X,Y$ stand for direct and crossed channel ($X,Y=C, D$) and ($i,j=1,2$), are given by
\be
%ec.(84) of the notes
\widetilde{\Gamma}_{\tau \pm}(XY^{*})_{ij} \equiv K_{\tau}^2 \; \frac{1}{2 M_{\tau}}
\int d_3 \; P_i(X) P_j(Y)^{*}  \slashed{L}_{\pm}^X \slashed{L}_{\pm}^{Y \dagger} \ ,
\label{bGXYM}
\ee
where 
\be
\label{ktau}
K_{\tau}^2=G_F^4 f_{M_1}^2 f_{M_2}^2   V_{M_1}^2 V_{M_2}^2\ .
\ee
From now on, we will pay our attention in a scenario where both mesons are equal, then $M_1 = M_2 \equiv M_M$ and the  constant $K_{\tau}^2 \equiv K_{M}^2$ presented in Eq.~(\ref{ktau}) becomes $K_{\pi}^2 =G_F^4 f_{\pi}^4 V_{u \bar{d}}^4 $ when the mesons are pions and $K_{K}^2 =G_F^4 f_{K}^4 V_{u \bar{s}}^4 $ when they are kaons. The canonical decay width has been evaluated numerically by means of \textit{Monte-Carlo} integrations using {\textit{Vegas}} algorithm \cite{Lepage:1980dq}\footnote{The integration were performed in two different languages \textit{Pyhton} and \textit{Fortran} in order to reduce the uncertainties.}. Furthermore, the evaluation were implemented using small  $\Gamma_{N_j} = 10^{-3}$ in the heavy neutrino propagators. The numerical results can be summarized as follows:
\begin{enumerate}[i)]
\item The contribution of $(DD^*)_{jj}$ and $(CC^*)_{jj}$ channels are approximately equal, thus  $\widetilde{\Gamma}_{\tau}(DD^{*})_{jj} \approx \widetilde{\Gamma}_{\tau}(CC^{*})_{jj}$.
\item The contribution of $(DC^*)_{ij}$ and $(CD^*)_{ij}$ channels are approximately equal, thus  $\widetilde{\Gamma}_{\tau}(DC^{*})_{ij} \approx \widetilde{\Gamma}_{\tau}(CD^{*})_{ij}$.
\item \label{ib} The terms $\widetilde{\Gamma}_{\tau}(DD^{*})_{jj} \propto 1/\Gamma_{N_j}$\footnote{It is important to note that the dependence $\widetilde{\Gamma}_{\tau}(DD^{*})_{jj} \propto 1/\Gamma_{N_j}$ is in agreement with the fact that sterile neutrino are weakly interacting particles and therefore the narrow width approximation $\frac{M_{N_j}}{(p_N^2-M_{N_j}^2)^2+(M_{N_j}\Gamma_{N_j})^2}\ \to \ \frac{ \pi}{\Gamma_{N_j}}\; \delta(p_{N}^2-M_{N_j}^2)$ is valid.}, while
 $\widetilde{\Gamma}_{\tau}(DC^{*})_{jj}$ and $\widetilde{\Gamma}_{\tau}(DC^{*})_{ij}$ are approximately independent of $\Gamma_{N_j}$.
\item When $\Gamma_{N_i} = 10^{-3}$,  the terms $ \widetilde{\Gamma}_{\tau \pm}(DC^{*})_{ii}$ and  $\widetilde{\Gamma}_{\tau \pm}(CD^{*})_{ii}$ are suppressed by a factor $\sim10^{-3}$, besides taking into account the latter point \ref{ib}, the terms $ \widetilde{\Gamma}_{\tau \pm}(DC^{*})_{jj}$ and  $\widetilde{\Gamma}_{\tau \pm}(DC^{*})_{ij}$
are negligible in all cases, in comparison with $\widetilde{\Gamma}_{\tau}(DD^{*})_{jj}$ and $\widetilde{\Gamma}_{\tau}(CC^{*})_{jj}$.
\item The contribution of $(DD^*)_{ij}$ and $(CC^*)_{ij}$ channels are approximately equal, and can reach the same order of magnitude than the $(DD^*)_{jj}$ and $(CC^*)_{jj}$ contributions\footnote{The effect of this kind of interference will be studied later in detail.}.
\end{enumerate}
Thus, under the above considerations and taking into account that $M_1=M_2=M_{\pi}, M_K$, we rewrite the Eq.~(\ref{dwp}) only in terms of the dominant contributions, as follows
\begin{subequations}
\label{tot}
\begin{align}
%ec.(83) of the notes
\label{dwft}
\Gamma(\tau^{\pm}) = &  \frac{1}{2!}
\sum_{i=1}^2 \sum_{j=1}^2 k_i^{(\pm)} k_j^{(\pm)*} \times {\big [}
\widetilde{\Gamma}_{\tau}(DD^{*})_{ij} + \widetilde{\Gamma}_{\tau}(CC^{*})_{ij} {\big ]}\\
\nonumber &= |B_{\ell N_1}^{}|^2 |B_{\tau N_1}^{}|^2 \widetilde{\Gamma}_{\tau}(DD^{*})_{11}+ |B_{\ell N_2}^{}|^2 |B_{\tau N_2}^{}|^2\widetilde{\Gamma}_{\tau}(DD^{*})_{22}
\nonumber \\
\nonumber
&+2 |B_{\ell N_1}^{}| |B_{\ell N_2}^{}| |B_{\tau N_1}^{}| |B_{\tau N_2}^{}| \widetilde{\Gamma}_{\tau}(DD^{*})_{11}  \cos(\theta_{12}) \delta_{12} \ , \\
& \mp 2 |B_{\ell N_1}^{}| |B_{\ell N_2}^{}| |B_{\tau N_1}^{}| |B_{\tau N_2}^{}| \widetilde{\Gamma}_{\tau}(DD^{*})_{11} \frac{\eta(y)}{y} \sin (\theta_{12}) \ , 
\label{dwf}
\end{align}
\end{subequations}

here $\delta_{12} \equiv \frac{\Re \big[ \widetilde{\Gamma}_{\tau}(DD^{*})_{12} \big]}{ \widetilde{\Gamma}_{\tau}(DD^{*})_{11}}$ measures the effect of $N_1 - N_2$ overlap\footnote{$\Re$ stand for the real part.}, the factor $\frac{\eta(y)}{y}$ will be discussed later, however, their values are presented in Fig.~\ref{delta_eta} and  $\theta_{12}=\phi_{\ell N_1}-\phi_{\ell N_2}+\phi_{\tau N_2}-\phi_{\tau N_1}$.
The diagonal canonical decay widths, presented in Eq.~(\ref{dwf}), can be implemented by means of the narrow width approximation  
\be
\widetilde{\Gamma}_{\tau}(DD^{*})_{jj}=\frac{K_{M}^2}{128 \pi^2 M_{\tau}^3 M_{N_j} \Gamma_{N_j}} \; \times \lambda^{1/2} \Bigg( 1, \frac{M_{\ell}^2}{M_N^2},\frac{M_{M}^2}{M_N^2}  \Bigg) \times Z(M_{\tau},M_{N_j},M_{M},M_{\ell}) \ ,
\ee
where the functions $Z(a, b, c, d)$ and $\lambda ( x, y, z) $ are kinematical functions, which are defined in Appendix \ref{PS}. 
The branching ratio for the process $\tau^{\pm} \to M_1^{\pm} M_2^{\pm} \ell^{\mp}$ is
\be
Br(\tau^{\pm})= \frac{\Gamma(\tau^{\pm})}{\Gamma(\tau^{\pm} \to \rm{all})} \ ,
\ee
where $\Gamma(\tau^{\pm} \to \rm{all})$ is the total decay width for $\tau^{\pm}$ lepton and is given by
\be
\Gamma(\tau^{\pm} \to \rm{all})=\frac{G_F^2 M_{\tau}^5}{192 \pi^3} \ .
\ee
In order to have a more realistic discussion, we must consider the acceptance factor, which is defined as the probability of the neutrino $N_j$ decay inside of a detector of length $L$
\begin{equation}
P_{N_j} \approx \frac{L}{\gamma_{N_j} \tau_{N_j} \beta_{N_j}} \approx \frac{L\, \Gamma_{N_j}}{\gamma_{N_j} \beta_{N_j }}
\label{PN}
\end{equation}
where $\gamma_{N_j}$ is the Lorentz time dilation factor in the Laboratory frame and $\beta$ is the neutrino speed\footnote{In this work, we will provide $\gamma_{N_j} \sim 2$, $\beta \sim 1$ and $L=1$ mts.}.  Therefore, the effective branching ratio\footnote{The $Br^{\txty{eff}}(\tau^{\pm})$ correspond to the real branching ratio, while $\Gamma^{\rm eff}(\tau^{\pm})$ correspond to the effective decay with, whose can be measured in an experiment.} is
\begin{equation}
\begin{split}
\textrm{Br}^{\txty{eff}}(\tau^{\pm})=P_{N_j} \textrm{Br}(\tau^{\pm})&=\frac{\Gamma^{\rm eff}(\tau^{\pm})}{\Gamma(\tau^{\pm} \to \rm{all})}=P_{N_j} \frac{\Gamma(\tau^{\pm})}{\Gamma(\tau^{\pm} \to \rm{all})}\,.
\end{split}
\label{BR-eff}
\end{equation}

%%%%%%%%%%%%%%%%%%%%%%%%%%%%%%%%%%%%%%%%%%%%%%%%
%%%%%%%%%%%%%%%%%%%%%%%%%%%%%%%%%%%%%%%%%%%%%%%%
%%%%%%%%%%%%%%%%%%%%%%%%%%%%%%%%%%%%%%%%%%%%%%%%
%%%%%%%%%%%%%%%%%%%%%%%%%%%%%%%%%%%%%%%%%%%%%%%%

\section{CP Asymmetry of $\tau^{\pm} \to M_1^{\pm} N_j \to M_1^{\pm} M_2^{\pm} \ell^{\mp}$ decays}
\label{sCP}
\begin{figure}[H]
\includegraphics[scale = 0.40]{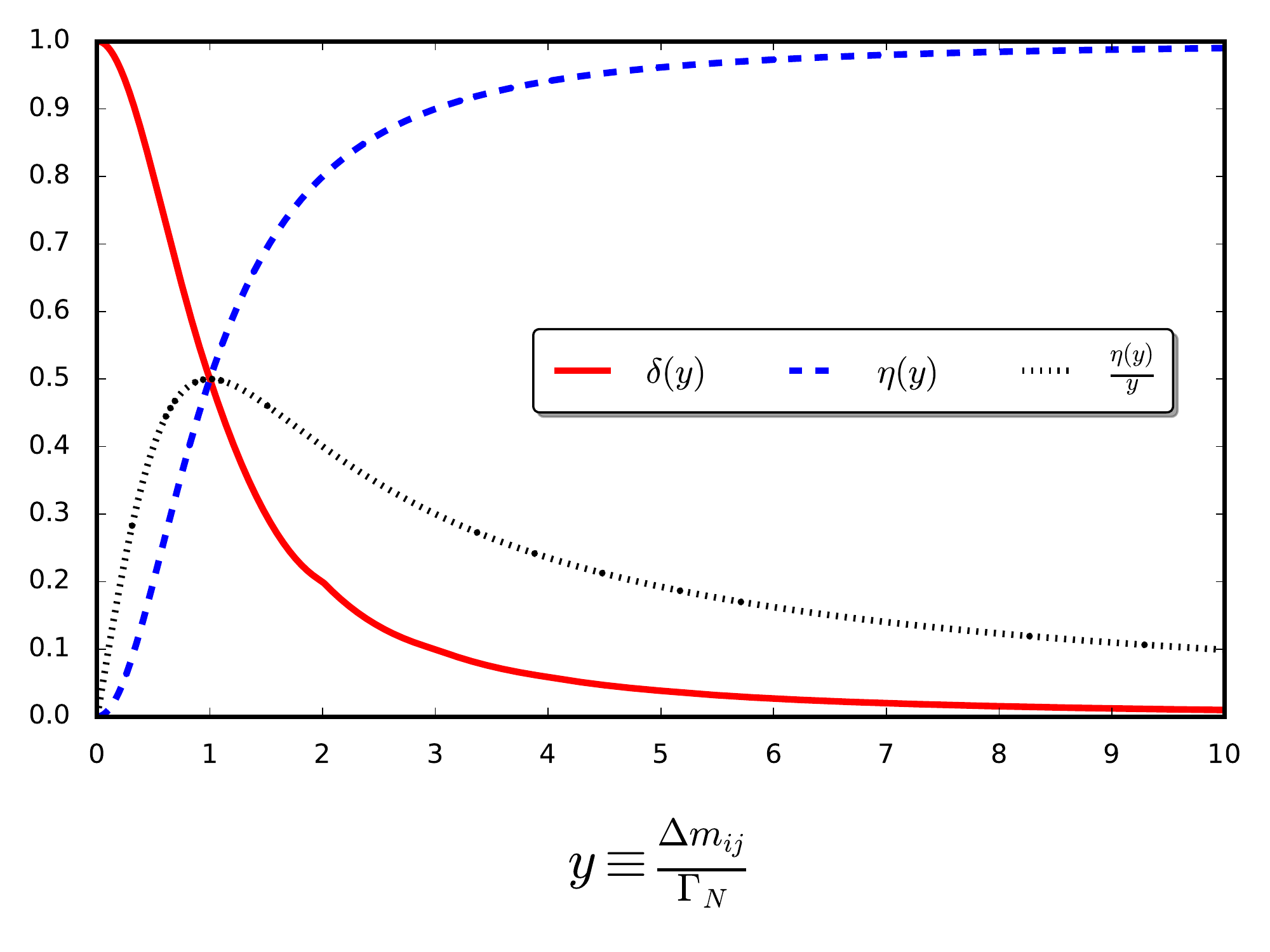}
\caption{Solid line (online red) overlap function $\delta_{12}$. Dashed line (online blue) $\eta(y)$ function. Dotted line (online black) $\eta(y)/y$ function.}
\label{delta_eta}
\end{figure}

In this section we will calculate the size of CP asymmetry $A_{CP}$, which is defined as follows
\begin{align}
\label{cpasym}
A_{CP}=\frac{\Gamma(\tau^{+})-\Gamma(\tau^{-})}{\Gamma(\tau^{+})+\Gamma(\tau^{-})} \ ,
\end{align}
The CP violation comes from the complex phases in the transition amplitudes Eq.~(\ref{amp}), and the observable effects only arise due to interference of at least two amplitudes. The CP-odd phases are those that come from the Lagrangian of the theory, in other words from the heavy-light mixing elements ($B_{\ell N}$);  these phases change sign between a process and its conjugate. On the other hand, the CP-even phases appear as absorptive parts in the propagators Eq.~(\ref{propa}) and do not change sign for the conjugate process. In order to have a more phenomenological discussion about $CP$ violation, it is useful define a new quantity  $A_{CP} Br^{\txty{eff}}(\tau^+)$ which is the corresponding branching ratio for the CP-violating asymmetry\footnote{In Eq.~(\ref{AcpEff}) we have used $\Gamma(\tau^{+})+\Gamma(\tau^{-})\approx 2 \Gamma(\tau^{+})$.}
\be
\label{AcpEff}
A_{CP}\  Br^{\txty{eff}}(\tau^+)=\frac{\Gamma(\tau^{+})-\Gamma(\tau^{-})}{\Gamma(\tau^{+})+\Gamma(\tau^{-})}\; Br^{\txty{eff}}(\tau^+)\approx P_{N_j}\; \frac{\Gamma(\tau^{+})-\Gamma(\tau^{-})}{2 \Gamma(\tau^{+} \to all)}
\ee
The CP-violating difference  $\Gamma(\tau^{+})-\Gamma(\tau^{-})$ is proportional to the imaginary part of $\widetilde{\Gamma}_{\tau}(DD^{*})_{12}$ and can be written as\footnote{Here we assumed the fact that $\Im \Big[ \widetilde{\Gamma}_{\tau}(DD^{*})_{12} \Big] \approx \Im \Big[ \widetilde{\Gamma}_{\tau}(CC^{*})_{12} \Big]$. }
\be
\label{asym}
\Gamma(\tau^{+})-\Gamma(\tau^{-}) \approx 4 |B_{\ell N_1}|  |B_{\ell N_2}|  |B_{\tau N_1}|  |B_{\tau N_2}| \ \sin{\theta_{12}}\  \Im \Big[ \widetilde{\Gamma}_{\tau}(DD^{*})_{12} \Big]
\ee
where we have neglected all the ($DC^*$) and ($CD^*$) interference contributions , due to fact that numerical simulation shows that they are strongly suppressed in comparison with ($DD^*$) and ($CC^*$). The imaginary part of Eq.~(\ref{asym}) correspond to the imaginary part of the off-diagonal elements in Eq.~(\ref{tot}) 
\be
\label{imparte}
\Im \Big[ \widetilde{\Gamma}_{\tau}(DD^{*})_{12} \Big]=\frac{1}{2 M_{\tau}} \int d_3 \; \Im \Big[P_1(D) P_2(D)^{*}\Big]  |\slashed{L}_{+}^D|^2 \ .
\ee
The imaginary part of the product of propagators (see Eq.~(\ref{impart}) in Appendix.~\ref{KR}) can be expressed using the narrow width approximation as
\bes
\label{ImP1P2gen}
\ba
{\rm Im} \left( P_1(D) P_2(D)^{*} \right)
&= &
\frac{
\left( p_N^2 - M_{N_1}^2 \right)  \Gamma_{N_2} M_{N_2}
- \Gamma_{N_1} M_{N_1} \left( p_N^2 - M_{N_2}^2 \right)
}
{
\left[ \left( p_N^2 - M_{N_1}^2 \right)^2 + \Gamma_{N_1}^2 M_{N_1}^2
\right]
\left[ \left( p_N^2 - M_{N_2}^2 \right)^2 + \Gamma_{N_2}^2 M_{N_2}^2
\right]
}
\label{ImP1P2exP}
\\
& \approx & \frac{\pi}{M^{2}_{N_2}-M^{2}_{N_1}} \left [
\delta  ( p^{2}_{N}-M^{2}_{N_2})+ \delta  ( p^{2}_{N}-M^{2}_{N_1})  \right ] \ ;
\label{ImP1P2a}
\ea
\ees
%where $p_N=(p_{\tau}-p_1)$ in the direct channel. 
the validity of Eq.~(\ref{ImP1P2a}) strongly depends on the assumption $\Gamma_{N_j} \ll | \Delta M_N | \equiv M_{N_2}-M_{N_1}$. However, it is useful introduce the parameter $\eta (y)$ where $y \equiv \frac{\Delta M_N}{\Gamma_N}=\frac{\Delta M_N}{\frac{1}{2}(\Gamma_{N_1}+\Gamma_{N_2})}$, which  parametrizes any deviation of Eq.~(\ref{ImP1P2exP}) when $\Gamma_{N_j} \not\ll | \Delta M_N |$ 
\be
\label{eta}
\eta (y)=\frac{\Im \Big[ \widetilde{\Gamma}_{\tau}(DD^{*})_{12} \Big]_{\textrm{NWA}}}{\Im \Big[ \widetilde{\Gamma}_{\tau}(DD^{*})_{12} \Big]_{\textrm{NUM}}}
\ee
In Eq.~(\ref{eta}) the subscripts   $NWA$ and $NUM$ stand for ''Narrow Width Approximation'' and ''Numerical'', respectively. The values of $\eta(y)$ were evaluated numerically using finite $\Delta M_N$ and their values are presented in Fig.~\ref{delta_eta} as a function of $y \equiv \Delta M_N/\Gamma_N$. The general expression of Eq.~(\ref{imparte}) including the $\eta(y)$ parameter and under the assumptions $M_{N_1}+M_{N_2} \approx 2 M_N$ is given by\footnote{Due to the fact that $\Gamma_N \sim \K_j^{Ma} \sim |B_{\ell N}|^2$ the mass difference becomes $\Delta M_N \ll 1$, hence the assumption $M_{N_1}+M_{N_2} \approx 2 M_N$ is reasonable In Eq.~(\ref{imeta}).}

\be
\label{imeta}
\Im \Big[ \widetilde{\Gamma}_{\tau}(DD^{*})_{12} \Big] \approx \eta(y)\; \frac{K_{M}^2}{128 \pi^2 M_{\tau}^3 M_{N} \Delta M_N } \; \times \lambda^{1/2} \Bigg( 1, \frac{M_{\ell}^2}{M_N^2},\frac{M_{M}^2}{M_N^2}  \Bigg) \times Z(M_{\tau},M_{N},M_{M},M_{\ell})
\ee
finally, the CP-violating difference becomes
\begin{align}
\label{asymf}
\nonumber
\Gamma(\tau^{+})-\Gamma(\tau^{-}) \approx & \eta(y)\; \frac{K_{M}^2 |B_{\ell N_1}|  |B_{\ell N_2}|  |B_{\tau N_1}|  |B_{\tau N_2}|}{32 \pi^2 M_{\tau}^3 M_{N} \Delta M_N } \; \sin{\theta_{12}} \\
& \times \lambda^{1/2} \Bigg( 1, \frac{M_{\ell}^2}{M_N^2},\frac{M_{M}^2}{M_N^2}  \Bigg) \times Z(M_{\tau},M_{N},M_{M},M_{\ell})
\end{align}
From Eq.~(\ref{cpasym}), Eq.~(\ref{asymf}) and Fig.~\ref{delta_eta} we can conclude that the best scenario for simultaneous maximization of $A{CP}$ and $Br(\tau)$, occurs when $y=1$. From now on, we will focus in a scenario where  heavy neutrinos are almost degenerate $\Delta M_N \sim \Gamma_N$; within this context we have assumed $|B_{\ell N_1}| \approx |B_{\ell N_2}| \equiv |B_{\ell N}|$, where $\ell = e, \mu, \tau$ and the mixing elements are $\K_1^{Ma} \approx  \K_2^{Ma} \equiv  \K^{Ma}$, therefore, the $CP$ asymmetry becomes
\be
A_{CP}\approx \eta(y)\; \frac{\Gamma_N}{\Delta M_N} \; \frac{\sin \theta_{12}}{1+\delta_{12} \cos \theta_{12}} \equiv  \frac{\eta(y)}{y} \; \frac{\sin \theta_{12}}{1+\delta_{12} \cos \theta_{12}} \ ,
\ee
consequently
\begin{align}
\label{effACP}
\nonumber
A_{CP}\  Br^{\txty{eff}}(\tau^+) \approx & \frac{\eta(y)}{y}\; \frac{L}{\gamma_N} \; |B_{\ell N}|^2 |B_{\tau N}|^2 \sin \theta_{12} 
\; \frac{3 \pi K_M^2}{2 G_F^2 M_{\tau}^8 M_N} \\
& \times  \lambda^{1/2} \Bigg( 1, \frac{M_{\ell}^2}{M_N^2},\frac{M_{M}^2}{M_N^2}  \Bigg) \times Z(M_{\tau},M_{N},M_{M},M_{\ell}) \ .
\end{align}

%%%%%%%%%%%%%%%%%%%%%%%%%%%%%%%%%%%%%%%%%%%%%%%%%%%%%%
%%%%%%%%%%%%%%%%%%%%%%%%%%%%%%%%%%%%%%%%%%%%%%%%%%%%%%

There is just one caveat in the expressions above: we have disregarded the effect of $N_1-N_2$ oscillation, these type of oscillations have been studied in detail in Ref.~\cite{Cvetic:2015ura} and it is straightforward to show that the $L$ dependent effective differential decay width is\footnote{In Eq.~(\ref{DWL}) $L$ is the distance between production vertex and detector; the quantities $\gamma_N$ and $\beta_N$ are: $\gamma_N=\frac{1}{2}(\gamma_{N_1} + \gamma_{N_2})$ and $\beta_N+\frac{1}{2}(\beta_{N_1} + \beta_{N_2})$, respectively.}
\begin{align}
\label{DWL}
\frac{d}{d L}& \Gamma_{\rm eff}^{\rm (osc)}(\tau^+ \to \pi^+ \pi^+ \mu^-;L) 
 \approx  \frac{1}{\gamma_N \beta_N}
\overline{\Gamma}(\tau^{+} \to \pi^{+} N) \overline{\Gamma}(N \to \pi^{+} \mu^{-})
\nonumber\\
& \times
\left\{ \sum_{j=1}^2 |B_{\mu N_j}|^2 |B_{\tau N_j}|^2 +
2  |B_{\mu N_1}| |B_{\tau N_1}| |B_{\mu N_2}| |B_{\tau N_2}|
\cos\left( L \frac{\Delta M_N}{\beta_N \gamma_N} + \theta_{12} \right)
\right\}
\end{align}
where $\overline{\Gamma}(\tau^{+} \to \pi^{+} N)$ and  $\overline{\Gamma}(N \to \pi^{+} \mu^{-})$ are kinematical functions presented in appendix \ref{KR}. In Eq.~(\ref{DWL}) it is also possible to notice that the oscillation length is $L_{\rm osc} = \frac{2 \pi \beta_N \gamma_N}{\Delta M_N}$. Then, the argument of cosine in Eq.~\ref{DWL} can be written as $2 \pi \frac{L}{L_{\rm osc}} + \theta_{12}$, therefore, in order to integrate out there are two possible scenarios:
\begin{enumerate}
\item $L \gg L_{\rm osc}$: In this regime we recover the main contributions of the  $L$-independent effective decay width (Eq.~(\ref{BR-eff})), because the oscillation term $\sim \cos \big(\mathpzc{f} (L)+  \theta_{12} \big)$ gives a relatively negligible contribution when integrated over several $L_{\rm osc}$. 
\item $L \nsim L_{\rm osc}$: In this scenario the integration of expression \ref{DWL} is
\begin{small}
\begin{align}
\label{WL}
\nonumber
\Gamma_{\rm eff}^{\rm (osc)}&(\tau^+ \to \pi^+ \pi^+ \mu^-;L) 
 \approx  \frac{L}{\gamma_N \beta_N}
\overline{\Gamma}(\tau^{+} \to \pi^{+} N) \overline{\Gamma}(N \to \pi^{+} \mu^{-}) \times
\Bigg[ \sum_{j=1}^2 |B_{\mu N_j}|^2 |B_{\tau N_j}|^2
\nonumber \\
&
+ \frac{L_{\rm osc}}{\pi L}  |B_{\mu N_1}| |B_{\tau N_1}| |B_{\mu N_2}| |B_{\tau N_2}|
\bigg( \sin \Big ( 2 \pi \frac{L}{L_{\rm osc}} + \theta_{12} \Big ) - \sin \big( \theta_{12} \big)\bigg)
\Bigg] \ , 
\end{align}
\end{small} in \ref{WL} we can see, immediately, that when $L_{\rm osc} \gg L$ and $L_{\rm osc} = L$   the oscillation effect disappear and we recover the $L$-independent main contributions of the Eq.~(\ref{BR-eff}). On the other hand, when $L \sim L_{\rm osc}$ neutrinos have traveled enough to have a well-defined oscillation, which means that neutrinos have not decayed yet (i.e.  $P_N \ll 1$). Moreover, $L \sim L_{\rm osc}$ means $y \equiv \frac{\Delta M_N}{\Gamma_N} \approx \frac{2 \pi}{P_N} \gg 1$ and then from Fig.~\ref{delta_eta} we notice that $y \gg 1$ destroy the effect of resonant CP violation. Therefore, the fact that disregard the $N_1 - N_2$ oscillation when we have chosen ${\eta(y)} \sim 1$ is valid. 
\end{enumerate} 

It is important to note that the oscillation effect is present when $L \sim L_{\rm osc}$, therefore, in general CP violating scenarios (i.e. when we are off CP resonant region) this must be taken into account.

%%%%%%%%%%%%%%%%%%%%%%%%%%%%%%%%%%%%%%%%%%%%%%%%%%%%%%
%%%%%%%%%%%%%%%%%%%%%%%%%%%%%%%%%%%%%%%%%%%%%%%%%%%%%%
%%%%%%%%%%%%%%%%%%%%%%%%%%%%%%%%%%%%%%%%%%%%%%%%%%%%%%
%%%%%%%%%%%%%%%%%%%%%%%%%%%%%%%%%%%%%%%%%%%%%%%%%%%%%%

\section{Results}
\label{sres}
\begin{figure}[h]
\includegraphics[scale = 0.36]{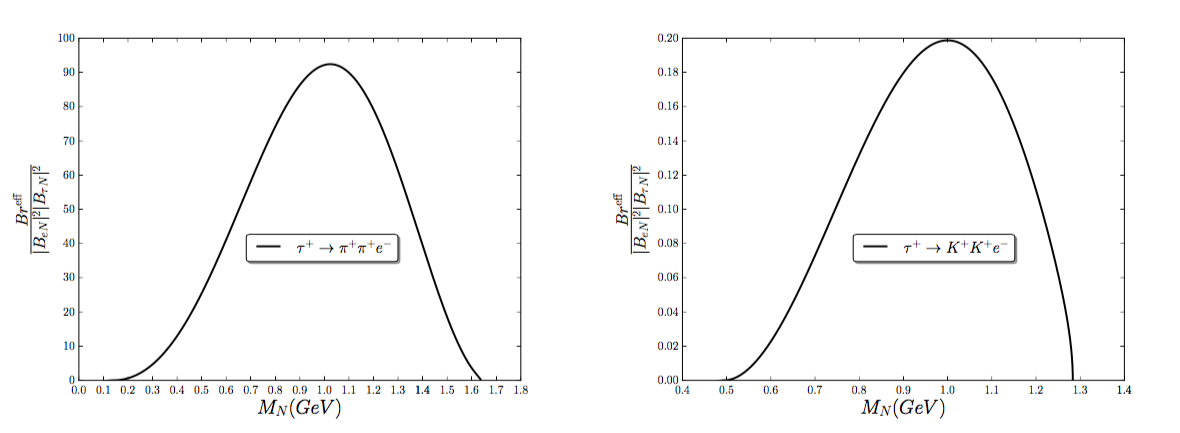}
\caption{Effective branching ratios per unit of $|B_{e N}|^2|B_{\tau N}|^2$. Here we use the following input parameters: $\cos \theta_{12} = 1/ \sqrt{2}$, overlap factor $\delta_{12}=0.5$, detector length $L=1$ mts, neutrino speed $\beta=1$ and Lorentz factor $\gamma_{N}=2$.}
\label{bre}
\end{figure}

\begin{figure}[h]
\includegraphics[scale = 0.36]{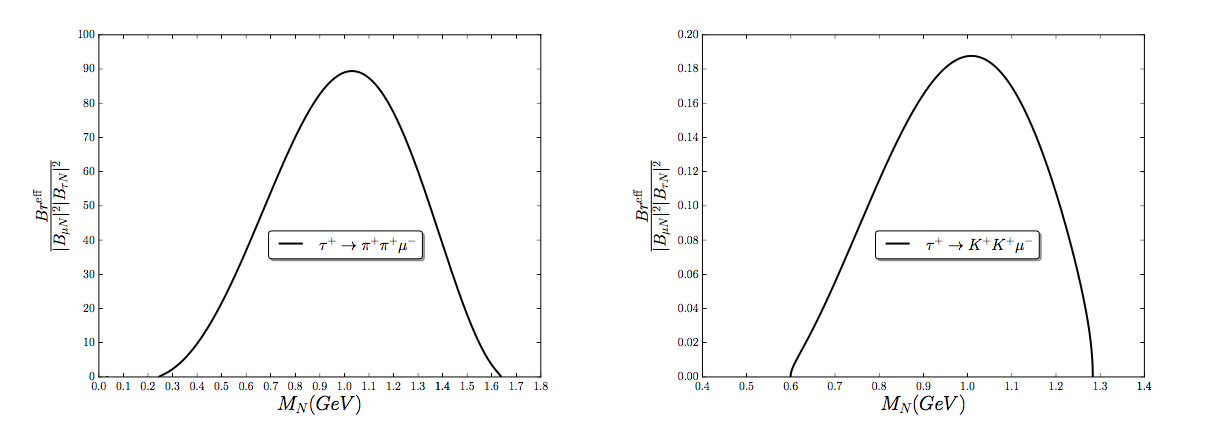}
\caption{Effective branching ratios per unit of $|B_{\mu N}|^2|B_{\tau N}|^2$. Here we use the following input parameters: $\cos \theta_{12} = 1/ \sqrt{2}$, overlap factor $\delta_{12}=0.5$, detector length $L=1$ mts, neutrino speed $\beta=1$ and Lorentz factor $\gamma_{N}=2$.}
\label{brmu}
\end{figure}

In this section the main results obtained in this work will be applied in order to provide a clue for future searches in tau factories.  The result for the effective branching ratios presented in Eq.~(\ref{BR-eff}) are shown in Fig.~\ref{bre} and Fig.~\ref{brmu} 

The  difference between the cases with $M_M = \pi$ and $M_M = K$ in the final states is mainly due to the elements of $CKM$ matrix, whereas for pions $V_{\pi} \approx 0.97$ and $V_{K} \approx 0.22$, respectively. Moreover,  the values of meson decay constant are $f_{\pi} \approx 0.13$ GeV and $f_{K} \approx 0.15$ GeV, therefore $K_{\pi}^2/K_{K}^2 \approx 2 \times 10^2$.
In order to estimate the region of heavy-light mixings elements $|B_{\ell N}|^2 |B_{\tau N}|^2$  which can be explored in future experiment\footnote{The Eq.~({\ref{explb}}) is presented in order to detect at least 1 event of difference between $Br(\tau^+)$ and $Br(\tau^-)$, here we have chosen $\eta(y)/y=1/2$.  } we define the following relation
\begin{align}
A_{CP} Br^{\txty{eff}}(\tau^+) \times N_{\tau} \ge 1 \quad \Rightarrow \quad |B_{\ell N}|^2 |B_{\tau N}|^2 \ge \frac{\gamma_N}{L N_{\tau}  \sin{\theta_{12}} \overline{S}(M_N) } \ ,
\label{explb}
\end{align}
here $N_{\tau}$ is the number of $\tau$ lepton produced in an experiment and $\overline{S}(M_N) $ is given by
\be
\overline{S}(M_N) = \frac{3 \pi K_{M}^2}{4 G_F M_{\tau}^8 M_N}\;  \lambda^{1/2} \Bigg( 1, \frac{M_{\ell}^2}{M_N^2},\frac{M_{M}^2}{M_N^2}  \Bigg) \times Z(M_{\tau},M_{N},M_{M},M_{\ell}) \ .
\ee
%%%%%%%%%%%%%%%%%%%%%%%%%%%%%%%%%%%%%%%%%%%%%%%%%%%%%%
%%%%%%%%%%%%%%%%%%%%%%%%%%%%%%%%%%%%%%%%%%%%%%%%%%%%%%
%%%%%%%%%%%%%%%%%%%%%%%%%%%%%%%%%%%%%%%%%%%%%%%%%%%%%%

\begin{figure}[h]
\includegraphics[scale = 0.27]{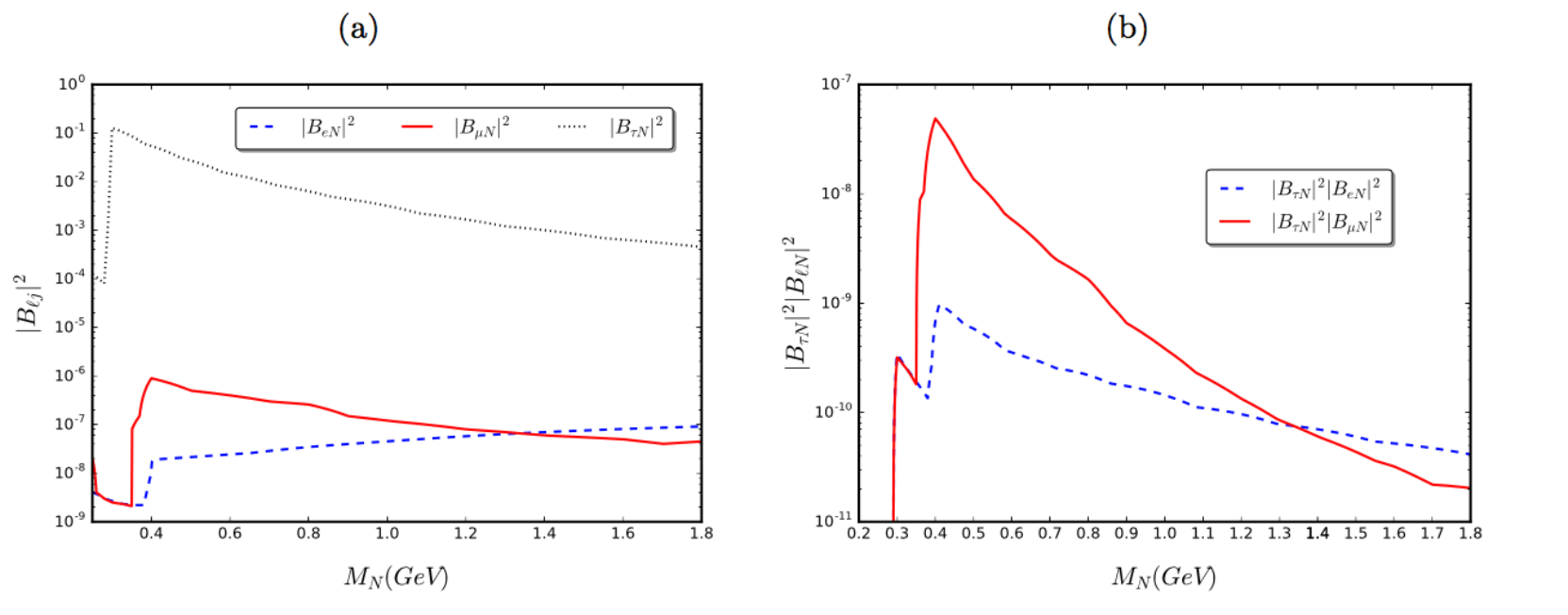}
\caption{(a) Exclusion regions of $|B_{\ell N}|^2$ taken from \cite{Atre:2009rg}. The dotted line (online black) stand for $|B_{\tau N}|^2$ , solid line (online red) stand for $|B_{\mu N}|^2$ and the dashed one (online blue) for $|B_{e N}|^2$. (b): Exclusion regions for the product of heavy-light mixings $|B_{\tau N}|^2|B_{\ell N}|^2$. The dashed line (online blue) stand for $|B_{\tau N}|^2 |B_{e N}|^2$ and the solid one (online red) for $|B_{\tau N}|^2 |B_{\mu N}|^2$. }
\label{explim}
\end{figure}

The actual experimental limits for heavy-light mixing elements are given in Ref.~\cite{Atre:2009rg}, and we have summarized them in Fig.~\ref{explim}(a) for the range of mass of interest. On the other hand, and due to the fact that our results depend on $|B_{\tau N}|^2 |B_{\ell N}|^2$, we present  in Fig.~\ref{explim}(b) the product of the experimental limits of interest.

\begin{figure}[H]
\includegraphics[scale = 0.32]{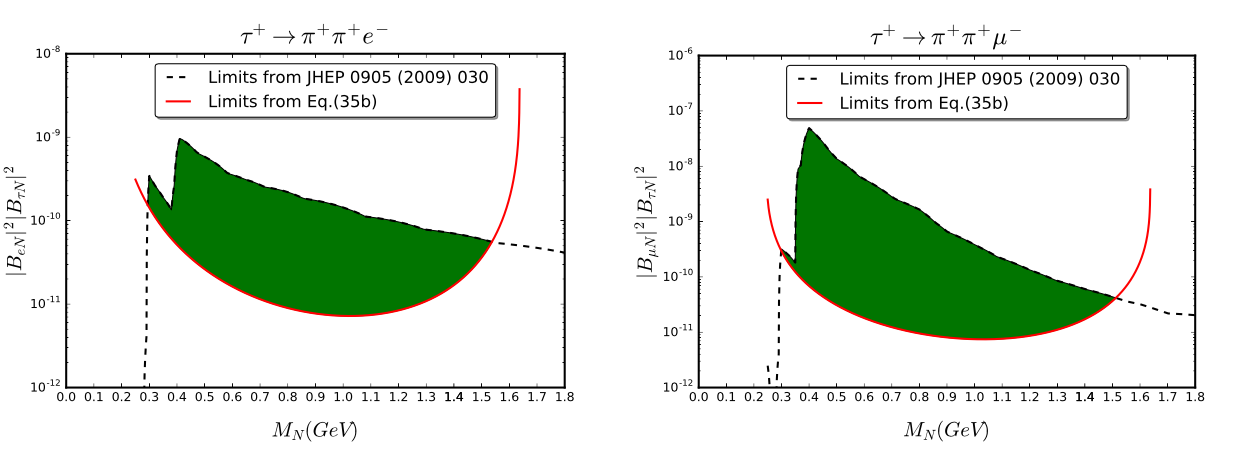}
\caption{The shaded region (online green) show the limits over the mixings parameter which could be reached in the future $\tau^{\pm}$ factory \cite{Eidelman:2015wja}. Right side: Limits for $|B_{e N}|^2|B_{\tau N}|^2$. Left side: Limits for $|B_{\mu N}|^2|B_{\tau N}|^2$. Here we use the following input parameters: $\eta(y)/y=0.5$, $N_{\tau}=10^{10}$,  $\cos \theta_{12} = 1/ \sqrt{2}$,   $L=1$ mts,  $\beta=1$ and  $\gamma_{N}=2$.}
\label{BRlim}
\end{figure}

The CTF in Novosibirsk, Russia is expected to collect $10^{10}$ pairs of $\tau^{\pm}$ leptons after few years of operation \cite{Eidelman:2015wja}, therefore under the latter considerations we can estimate the mixing region that can be explored in such experiment, this region is presented in Fig.~\ref{BRlim}.

It is important to point out that due to the $CKM$ elements suppresion only channels with pions in the final state offer real possibilities to constrain the heavy-light mixings parameters.

%%%%%%%%%%%%%%%%%%%%%%%%%%%%%%%%%%%%%%%%%%%%%%%%
%%%%%%%%%%%%%%%%%%%%%%%%%%%%%%%%%%%%%%%%%%%%%%%%
%%%%%%%%%%%%%%%%%%%%%%%%%%%%%%%%%%%%%%%%%%%%%%%%
%%%%%%%%%%%%%%%%%%%%%%%%%%%%%%%%%%%%%%%%%%%%%%%%

\section{Summary and Conclusions}
\label{s_conclusion}

In this letter we studied the ($\Delta L =2$) rare tau decays $\tau^{\pm} \to M_1^\pm  M_2^\pm \ell^{\mp}$, where $M_1$ and $M_2$ are pseudo scalar mesons ($M_1,\ M_2 = \pi, K$) and the charged lepton can be $\ell = e, \mu$, also we studied the possibility of $CP$ violation detection in future tau factories. We have assumed that the decays occur via the exchange of two on-shell sterile neutrinos $N_j$ at tree level, and we have shown that the amplitude of these processes is suppressed by the mixing elements of the PMNS matrix $|B_{\tau N}|^2 |B_{\ell N}|^2$. The aforementioned $CP$ violation  effects come from the interference between the $N_1$ and $N_2$ propagators and the complex phases (CP-odd phases $\phi_{\ell N_j}$, see Eq.~(\ref{PMNS})) in the PMNS mixing matrix. Our results shows that these signals of $CP$ violation could be detected in future tau factories for $\tau^{\pm} \to \pi^{\pm}  \pi^{\pm} \ell^{\mp}$ tau decays, where $\ell = e, \mu$ if there exist, at least, two sterile neutrinos in the on-shell mass range, their masses are almost degenerate $\Delta M_N \sim \Gamma_N$, the $CP$ odd phases $\sin \theta_{12} \not\ll 1$ and the mixing parameters are in the allowed region of Fig.~\ref{explim}. In such a case, the CP-violating difference $\Gamma({\tau^+})-\Gamma(\tau^-)$ becomes large and comparable with $\Gamma({\tau^+})+\Gamma(\tau^-)$ and the corresponding $CP$ asymmetry $A_{CP}$ becomes $ A_{CP} \sim 1$.  In addition, there exist several models with quasi-degeneracy $\Delta M_N \sim \Gamma_N$, between them it is worth to mention the well-know $\nu$MSM model \cite{Asaka:2005an,Asaka:2005pn}, where the quasi-degeneracy of the two heavy neutrinos (with mass $M_{N_j} \sim 1$ GeV) is fundamental in order to get a successful dark matter candidate. However, our results can be framed in the context of the $\nu$MSM model or more general models \cite{Drewes:2015iva,Zamora-Saa:2016qlk} with at least two quasi-degenerate neutrinos.
%%%%%%%%%%%%%%%%%%%%%%%%%%%%%%%%%%%%%%%%%%%%%%%%%%%%%%%
%%%%%%%%%%%%%%%%%%%%%%%%%%%%%%%%%%%%%%%%%%%%%%%%%%%%%%%

\section{Acknowledgments}
This work was supported by Fellowship Grant {\it Becas Chile} No.~74160012, CONICYT~(J.Z.S). Also, the author want to thank for valuable discussions with David Alvarez-Castillo.
%%%%%%%%%%%%%%%%%%%%%%%%%%%%%%%%%%%%%%%%%%%%%%%%
%%%%%%%%%%%%%%%%%%%%%%%%%%%%%%%%%%%%%%%%%%%%%%%%
%%%%%%%%%%%%%%%%%%%%%%%%%%%%%%%%%%%%%%%%%%%%%%%%
%%%%%%%%%%%%%%%%%%%%%%%%%%%%%%%%%%%%%%%%%%%%%%%%

\appendix
\section{Amplitude and kinematic relations for $\tau^{\pm} \to M_1^{\pm} M_2^{\pm} \ell^{\mp} $}
\label{KR}
The amplitude for the process via two on-shell intermediate heavy neutrino is
\begin{footnotesize}
\begin{align}
\label{ampsqp}
\nonumber \overline{|\mathcal{M_+}|^2}&= K_{\tau}^2 \Bigg[ |B_{\ell 1}^{}|^2 |B_{\tau 1}^{}|^2 \Big(|P_{1}(D)|^2\ |\slashed{L}_+^D|^2 + |P_{1}(C)|^2\ |\slashed{L}_+^C|^2  \Big) \\
\nonumber &+ |B_{\ell 2}^{}|^2 |B_{\tau 2}^{}|^2 \Big(|P_{2}(D)|^2\ |\slashed{L}_+^D|^2 + |P_{2}(C)|^2\ |\slashed{L}_+^C|^2  \Big)\\
\nonumber & + 2 |B_{\ell 1}^{}| |B_{\tau 1}^{}| |B_{\ell 2}^{}| |B_{\tau 2}^{}| \cos{\theta_{21}} \Big ( \Re \big[ P_{1}(D) P_{2}(D)^{*}  \big] |\slashed{L}_+^D|^2 +  \Re \big[ P_{1}(C) P_{2}(C)^{*}  \big]\ |\slashed{L}_+^C|^2 \Big)\\
\nonumber & + \Big ( 2  |B_{\ell 1}^{}|^2 |B_{\tau 1}^{}|^2 \Re \big[ P_{1}(D) P_{1}(C)^{*}  \big] +   2|B_{\ell 2}^{}|^2 |B_{\tau 2}^{}|^2 \Re \big[ P_{2}(D) P_{2}(C)^{*}  \big] \slashed{L}_+^{D} \slashed{L}_+^{C \dagger}\\
 \nonumber&+B_{\ell 1}^{} B_{\tau 1}^{*} B_{\ell 2}^{*} B_{\tau 2}^{} \Big( P_{1}(D) P_{2}(C)^{*}\ \slashed{L}_+^{D} \slashed{L}_+^{C \dagger} + P_{1}(C) P_{2}(D)^{*}\ \slashed{L}_+^{C} \slashed{L}_+^{D \dagger}  \Big ) \\&+B_{\ell 1}^{*} B_{\tau 1}^{} B_{\ell 2}^{} B_{\tau 2}^{*} \Big( P_{2}(D) P_{1}(C)^{*}\ \slashed{L}_+^{D} \slashed{L}_+^{C \dagger} + P_{2}(C) P_{1}(D)^{*}\ \slashed{L}_+^{C} \slashed{L}_+^{D \dagger}  \Big ) \Bigg]
\end{align}
\begin{align}
\label{ampsqm}
\nonumber \overline{|\mathcal{M_-}|^2}&= K_{\tau}^2 \Bigg[ |B_{\ell 1}^{}|^2 |B_{\tau 1}^{}|^2 \Big(|P_{1}(D)|^2\ |\slashed{L}_-^D|^2 + |P_{1}(C)|^2\ |\slashed{L}_-^C|^2  \Big) \\
\nonumber & + |B_{\ell 2}^{}|^2 |B_{\tau 2}^{}|^2 \Big(|P_{2}(D)|^2\ |\slashed{L}_-^D|^2 + |P_{2}(C)|^2\ |\slashed{L}_-^C|^2  \Big)\\
\nonumber & + 2 |B_{\ell 1}^{}| |B_{\tau 1}^{}| |B_{\ell 2}^{}| |B_{\tau 2}^{}| \cos{\theta_{21}} \Big ( \Re \big[ P_{1}(D) P_{2}(D)^{*}  \big] |\slashed{L}_-^D|^2 +  \Re \big[ P_{1}(C) P_{2}(C)^{*}  \big]\ |\slashed{L}_-^C|^2 \Big)\\
\nonumber & + \Big ( 2  |B_{\ell 1}^{}|^2 |B_{\tau 1}^{}|^2 \Re \big[ P_{1}(D) P_{1}(C)^{*}  \big] +   2|B_{\ell 2}^{}|^2 |B_{\tau 2}^{}|^2 \Re \big[ P_{2}(D) P_{2}(C)^{*}  \big] \slashed{L}_-^{D} \slashed{L}_-^{C \dagger}\\
 \nonumber&+B_{\ell 1}^{*} B_{\tau 1}^{} B_{\ell 2}^{} B_{\tau 2}^{*} \Big( P_{1}(D) P_{2}(C)^{*}\ \slashed{L}_-^{D} \slashed{L}_-^{C \dagger} + P_{1}(C) P_{2}(D)^{*}\ \slashed{L}_-^{C} \slashed{L}_-^{D \dagger}  \Big ) \\&+B_{\ell 1}^{} B_{\tau 1}^{*} B_{\ell 2}^{*} B_{\tau 2}^{} \Big( P_{2}(D) P_{1}(C)^{*}\ \slashed{L}_-^{D} \slashed{L}_-^{C \dagger} + P_{2}(C) P_{1}(D)^{*}\ \slashed{L}_-^{C} \slashed{L}_-^{D \dagger}  \Big ) \Bigg].
\end{align}
\end{footnotesize}\\

The kinematical factors presented in Eq.~(\ref{amp}), Eq.~(\ref{ampsqp}) and Eq.~(\ref{ampsqm}) are given by
\begin{footnotesize}
\begin{align}
\nonumber
|\slashed{L}_+^D|^2&=|\slashed{L}_-^D|^2=32(p_1 \cdot p_2)(p_2 \cdot p_{\ell})(p_1 \cdot p_{\tau})-16 M_2^2 (p_1 \cdot p_{\tau})(p_1 \cdot p_{\ell}) \\
 &-16 M_1^2 (p_2 \cdot p_{\tau})(p_2 \cdot p_{\ell})+8 M_1^2 M_2^2 (p_{\ell} \cdot p_{\tau})\\
 \nonumber \\
 \nonumber
|\slashed{L}_+^C|^2&=|\slashed{L}_-^C|^2=32(p_1 \cdot p_2)(p_1 \cdot p_{\ell})(p_2 \cdot p_{\tau})-16 M_1^2 (p_2 \cdot p_{\tau})(p_2 \cdot p_{\ell}) \\
 &-16 M_2^2 (p_1 \cdot p_{\tau})(p_1 \cdot p_{\ell})+8 M_1^2 M_2^2 (p_{\ell} \cdot p_{\tau})\\
 \nonumber \\
\nonumber 
\slashed{L}_{\pm}^D \slashed{L}_{\pm}^{C \dagger}&=\mp16 i \epsilon_{p1,p2,p_{\ell},p_{\tau}}(p_1 \cdot p_2)+16 M_2^2 (p_1 \cdot p_{\tau})(p_1 \cdot p_{\ell})+16 M_1^2 (p_2 \cdot p_{\tau})(p_2 \cdot p_{\ell})\\
&+16 (p_1 \cdot p_2)^2 (p_{\ell} \cdot p_{\tau}) -16(p_1 \cdot p_2)(p_2 \cdot p_{\ell})(p_1 \cdot p_{\tau})-16(p_1 \cdot p_2)(p_1 \cdot p_{\ell})(p_2 \cdot p_{\tau})\\
\nonumber \\
\slashed{L}_{\pm}^{D \dagger} \slashed{L}_{\pm}^{C}&=\Big( \slashed{L}_{\pm}^D \slashed{L}_{\pm}^{C \dagger} \Big )^*
\end{align}
\end{footnotesize}

The product of propagators $P_{1}(X) P_{2}(X)^*$ (where $X= D, C$) can be expressed as the sum of the real and imaginary parts
\begin{subequations}
\begin{small}
\begin{align}
\label{realpart}
P_{1}(X) P_{2}(X)^* &=  \underbrace{M_{N_1} M_{N_2} \frac{(P_{N}^{2}(X)-M_{N_1}^2)(P_{N}^{2}(X)-M_{N_2}^2)+\Gamma_{N_1} \Gamma_{N_2} M_{N_1}M_{N_2}}{\Big((P_{N}^{2}(X)-M_{N_1}^2)^2+ \Gamma_{N_1}^2 M_{N_1}^2\Big)\Big((P_{N}^{2}(X)-M_{N_2}^2)^2+ \Gamma_{N_2}^2 M_{N_2}^2\Big)}}_{\txty{Rear part}}\\
\nonumber \\
\label{impart}
& -i \; \underbrace{M_{N_1} M_{N_2} \frac{(P_{N}^{2}(X)-M_{N_2}^2)M_{N_1} \Gamma_{N_1} - (P_{N}^{2}(X)-M_{N_1}^2)M_{N_2} \Gamma_{N_2}}{\Big((P_{N}^{2}(X)-M_{N_1}^2)^2+ \Gamma_{N_1}^2 M_{N_1}^2\Big)\Big((P_{N}^{2}(X)-M_{N_2}^2)^2+ \Gamma_{N_2}^2 M_{N_2}^2\Big)}}_{\txty{Imaginary part}}
\end{align}
\end{small}
\end{subequations}

The partial decay widths presented in Eq.~(\ref{DWL}) are:
\begin{subequations}
\label{bG}
\bea
\overline{\Gamma}(\tau^{\pm} \to \pi^{\pm} N) & = &
\frac{1}{8 \pi} G_F^2 f_{\pi}^2 |V_{\pi}|^2 \frac{1}{M_{\tau}} 
\; \lambda^{1/2}\left(1, \frac{M_{\pi}^2}{M_{\tau}^2}, \frac{M_N^2}{M_{\tau}^2}\right)\times
\nonumber \\ 
&& \left[ \Big(M_{\tau}^2 - M_{N}^2\Big)^2 - M_{\pi}^2 \Big( M_{\tau}^2 + M_{N}^2 \Big) \right] \ ,
 \\
\overline{\Gamma}(N \to \mu^{\pm} \pi^{\mp}) & = &
\frac{1}{16 \pi} G_F^2 f_{\pi}^2 |V_{\pi}|^2 \frac{1}{M_N} 
\; \lambda^{1/2}\left(1, \frac{M_{\pi}^2}{M_N^2}, \frac{M_e^2}{M_N^2}\right)\times 
\nonumber \\ 
&& \left[ \Big(M_N^2 + M_e^2\Big) \Big(M_N^2 - M_{\pi}^2+M_e^2\Big) -
4 M_N^2 M_{e}^2 \right]\ .
\eea
\end{subequations}

%%%%%%%%%%%%%%%%%%%%%%%%%%%%%%%%%%%%%%%%%%%%%%%%
%%%%%%%%%%%%%%%%%%%%%%%%%%%%%%%%%%%%%%%%%%%%%%%%
%%%%%%%%%%%%%%%%%%%%%%%%%%%%%%%%%%%%%%%%%%%%%%%%
%%%%%%%%%%%%%%%%%%%%%%%%%%%%%%%%%%%%%%%%%%%%%%%%

\section{Phase space relations }
\label{PS}
The integration presented in Eq.~(\ref{dw}) can be performed in the following way:
\begin{align}
\label{dwE}
 \Gamma(\tau^{\pm}) &=  \frac{1}{2!}(2 - \delta_{M_1 M_2})\frac{1}{64 \pi^3 M_{\tau}}\int \overline{|\mathcal{M}_{\pm}|^2}\ dE_1\ dE_2\ ;
\end{align}
the integration limits over $E_2$ and $E_1$ for the ($DD^*$) channel are
\begin{align}
&E_2 \geq\frac{1}{2 m_{23}^2}\Bigg( (M_{\tau}-E_1)(m_{23}^2+M_2^2-M_3^2) - \sqrt{(E_1^2-M_1^2)\lambda(m_{23}^2,M_2^2,M_3^2)} \Bigg )\ , \\
\nonumber \\
&E_2 \leq\frac{1}{2 m_{23}^2}\Bigg( (M_{\tau}-E_1)(m_{23}^2+M_2^2-M_3^2) + \sqrt{(E_1^2-M_1^2)\lambda(m_{23}^2,M_2^2,M_3^2)} \Bigg ) \ , \\
\nonumber \\
& \quad \quad \quad \quad \quad \quad \quad \quad M_1 \leq E_1 \leq \frac{M_{\tau}^2+M_1^2-(M_2+M_3)^2 }{2 M_{\tau}} \ , 
\end{align}
where 
\be
m_{23}^2=M_{\tau}^2+M_1^2-2 M_{\tau}E_1 \ .
\ee
Finally, the kinematical functions $\lambda(x,y,z)$ and $Z(a,b,c,d)$ are

\be
\lambda(x,y,z)=x^2 + y^2 + z^2 -2 xy -2 xz -2 yz
\ee
\begin{align}
\nonumber
Z(a,b,c,d)= &\Big( (b^2-d^2)^2-c^2(d^2+b^2) \Big) \Big(  (a^2-b^2)^2-c^2(b^2+a^2 \Big) \\
& \times \sqrt{\Big( a^2 -(b-c)^2  \Big) \Big( a^2 -(b+c)^2  \Big)} 
\end{align}

%%%%%%%%%%%%%%%%%%%%%%%%%%%%%%%%%%%%%%%%%%%%%%%%
%%%%%%%%%%%%%%%%%%%%%%%%%%%%%%%%%%%%%%%%%%%%%%%%
%%%%%%%%%%%%%%%%%%%%%%%%%%%%%%%%%%%%%%%%%%%%%%%%
%%%%%%%%%%%%%%%%%%%%%%%%%%%%%%%%%%%%%%%%%%%%%%%%

\newpage

\bibliographystyle{apsrev4-1}

\bibliography{biblio.bib}

\end{document}

%% file: def_jazs_tau.tex
\usepackage{amsmath,amsfonts,amssymb,amsthm,bm,bbm,bbold,braket,color,dsfont,fancybox,hyperref,
srcltx,slashed,ucs,yfonts,cancel,xfrac,verbatim,xcolor}
\usepackage[FIGTOPCAP]{subfigure}
\usepackage{slashed}
\usepackage{floatrow}
\usepackage{graphicx}
\usepackage[outercaption]{sidecap} 

\usepackage[inline,shortlabels]{enumitem}
\usepackage{nicefrac}
\usepackage{multirow}

\DeclareMathAlphabet{\mathpzc}{OT1}{pzc}{m}{it}
\definecolor{gold}{rgb}{1,0.8,0}
\definecolor{nara}{rgb}{1,0.4,0.1}
\definecolor{goldo}{rgb}{1,0.7,0}
\definecolor{greeno}{rgb}{0,0.8,0}
\def\bes{\begin{subequations}}
\def\ees{\end{subequations}}
\def\be{\begin{equation}}
\def\ee{\end{equation}}
\def\bea{\begin{eqnarray}}
\def\eea{\end{eqnarray}}
\def\ba{\begin{eqnarray}}
\def\ea{\end{eqnarray}}
\def\bear{\begin{array}}
\def\eear{\end{array}}

\newcommand{\txty}[1]{\textrm{\tiny{#1}}}

\newcommand{\bpm}{\begin{pmatrix}}
\newcommand{\epm}{\end{pmatrix}}
\newcommand{\BM}{\left(\begin{array}}		% matrices
\newcommand{\BMC}{\left[\begin{array}}		% matrices
\newcommand{\EM}{\end{array}\right)}		% matrices
\newcommand{\EMC}{\end{array}\right]}		% matrices
	% ...
	% ...
 % ...

\newcommand{\com}[1]{\begin{comment}\end{comment}}
\newcommand{\K}{\mathcal{K}}